# Ultrabroadband, high color purity multispectral color filter arrays

Jiewei Xiang, Meiting Song, Yi Zhang, Jennifer Kruschwitz, Jaime Cardenas[*]

The Institute of Optics, University of Rochester, Rochester, N.Y. 14627, USA

The demand for multispectral imagers is on the rise in various fields, such as remote sensing, medical imaging, and material analysis, due to their ability to capture and analyze scenes or objects across multiple wavelength bands. This enables the extraction of valuable information related to the object's composition, reflectance properties, and other characteristics that are not distinguishable to RGB cameras. [1–3] However, conventional multispectral imagers are typically bulky and slow, requiring mechanical motion for spatial or spectral scanning. A solution to this limitation is the development of fast snapshot multispectral imaging systems that eliminate the need for scanning by incorporating multispectral filter arrays (MSFAs) which directly divide incoming light into multiple wavelength bands onto sensors or cameras.[3] To achieve the level of integration and compactness in modern digital cameras, it is essential to have high-performance and miniature MSFAs that can be directly integrated or fabricated onto CMOS sensors whose pixel sizes are on the order of a few micrometers. Recent advancements in nanofabrication technologies have demonstrated the feasibility of miniature MSFAs with individual filter sizes under 10 micrometers, utilizing thin-film stacks[4–11] or metallic and dielectric metasurfaces. [12–28] However, most designs involve trade-offs between different essential aspects of multispectral imaging, including high transmission, narrow linewidth (high spectral resolution), color purity (low sidebands), a large working wavelength range, angular and polarization insensitivity, and compatibility with current CMOS fabrication processes. Combining all these features into MSFAs with dimensions in the tens of micrometers poses a significant challenge.

Current multispectral filter arrays must trade-off transmission, bandwidth, angular sensitivity, and CMOS compatibility and often only excel in a few of them while compromising the others. Most miniature multispectral filter arrays (MSFAs) utilize different mechanisms of resonances to filter the incident light such as plasmonic resonances, Mie resonances, guide-mode resonances, cavity resonances, etc. Metallic metasurface-based MSFAs that use plasmonic resonances exhibit wide working wavelength ranges and simplicity in fabrication, but the transmission is limited by metallic absorption.[20–28] On the other hand, color filters that use dielectric metasurfaces with Mie resonances overcome this limitation but lack spectral selectivity, making it challenging to achieve high color purity and narrow linewidth.[16–18] Guided-mode resonance color filters that combine diffraction gratings and waveguide resonances can produce narrow linewidths and high transmission.[12,14,15,29] However, they are highly sensitive to incident angles due to their diffractive nature. Conventional thin film Fabry–Pérot color filters are relatively insensitive to polarization and angular changes as their resonances are not reliant on diffraction structures. They consist of a low-index cavity layer sandwiched between two distributed Bragg mirrors or metallic mirrors to filter incident light based on interference-induced cavity resonances.[4,9–11,30–32] In MSFA based on Fabry–Pérot structures, the peak wavelength of each color filter can be tuned by changing the optical length of the corresponding cavity layer. This is traditionally done by directly changing the physical thickness of the cavity, which may require multiple lithography steps[4,8–11] or special lithography technologies.[5–7] Therefore, it is challenging to fabricate more than 20 channels due to the increasing number of fabrication steps and the difficulty in integration of a staircase surface. To overcome these challenges, subwavelength structures have been introduced into the cavity layer, allowing peak wavelength tuning by changing the fill ratio, which leads to an effective index change of the cavity layer.[30–32] However, due to a limited free spectral range (FSR), Fabry–Pérot based MSFAs are restricted to a narrow working wavelength range to maintain high color purity.

In order to break the performance trade-off and achieve a CMOS compatible, miniature, broadband, multispectral filter arrays, MSFAs, with narrow linewidth, high transmission and color purity, we propose and demonstrate a modified, high-order Fabry–Pérot MSFA with selective peak suppression. We use subwavelength structures to tune the filter without changing the physical thickness and an ultrathin metal layer in the cavity to utilize high-order resonances ($2^{nd}$) (Figure 1a). This strategy breaks the trade-off between linewidth and FSR in a Fabry–Pérot resonator and exploits the reflection phase alleviation for high-order resonances. We use polysilicon and silicon dioxide to form two high reflection Bragg mirrors, which leads to a high transmission and narrow linewidth. By inserting a platinum layer in the middle of the cavity, we selectively suppress the odd-order resonances (mainly first and third-order resonances). Therefore, we can use the second-order resonance, which has a narrower linewidth than the widely adopted first-order resonance,[6,9,10] with an extended FSR and working wavelength range (Figure 1c (i-ii)). Furthermore, the fourth and higher-order resonances are suppressed by polysilicon absorption. A high color purity across the wavelength range from 400nm to 1000nm is then obtained with all the other resonances suppressed (Figure 1c (iii)). The cavity contains mesh or grating subwavelength structures to maximize the tuning range with reasonable fabrication challenges (Figure 1b). Without introducing high-order diffraction, our design inherits the relative insensitivity of polarization and angle from thin film Fabry–Pérot filters. All materials and fabrication processes are CMOS compatible to make it ready to integrate with current CMOS sensors. In addition to the second-order resonance, we can further expand the wavelength range by adjusting the position of the metal layer to incorporate the third-order resonance. By combining the second and third-order resonance designs, our MSFAs can cover the entire detection range of most silicon photodetectors.

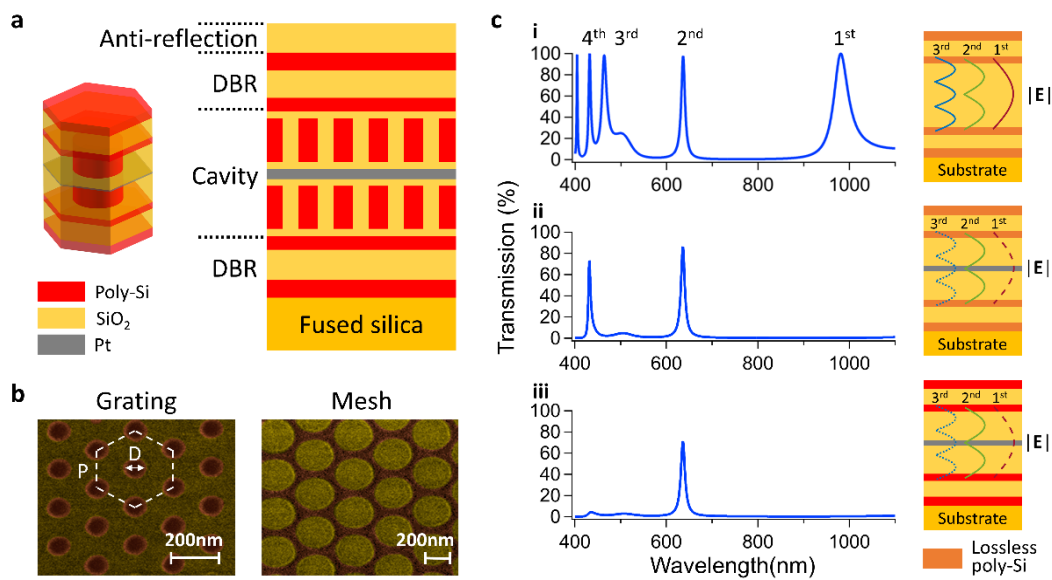

Figure 1. (a) Schematic diagram illustrating the proposed modified Fabry–Pérot color filter. The color filter is composed of HLH (high-low-high) thin film stacks consisting of polysilicon and $SiO_2$ forming

distributed Bragg reflectors (DBRs). A 10nm Pt layer separates the cavity, and the bandpass of each filter is determined by the mixture of polysilicon and SiO$_2$ in the cavity (see measured refractive index of used materials in Figure S1 in Supplementary information). (b) Top-view SEM images (false coloring highlights polysilicon (red) and silicon dioxide (yellow)) showcasing two different cavity structures. The grating structure comprises a hexagonal array of polysilicon nanoposts surrounded by SiO$_2$, while the mesh structure features a hexagonal array of SiO$_2$ nanoposts surrounded by polysilicon. (c) Illustration of selective suppression using an un-patterned cavity. (i) Simulated transmission spectrum of the proposed design without metal in the cavity and neglecting absorption from polysilicon. (ii) Addition of the metal layer in the cavity. (iii) Consideration of the absorption from polysilicon.

One of the major limitations of Fabry–Pérot based MSFAs lies in the trade-off between the FSR and the full width half maximum (FWHM), which is affected by the choice of the resonance order according to the resonance condition. A general Fabry–Pérot cavity has a resonance at every wavelength that satisfies the resonance condition

$$\lambda_m = \frac{4\pi n d \cos\theta'}{2m\pi - \phi_r^{total}},$$

where $m$ is the order of resonance, $\lambda_m$ is the $m^{th}$ order resonant wavelength, $n$ is the refractive index of the cavity material, $\theta$ is the refracted angle, $\phi_r^{total}$ is the sum of the reflection phases from the two mirrors. As a result, the working wavelength range of Fabry–Pérot based MSFAs is largely affected by the FSR of the resonant peaks, since multiple transmission peaks in the detection wavelength range can largely decrease the color purity. For a larger FSR, it is preferable to use lower-order resonances, such as the first-order resonant peak, since the FSR decreases rapidly for higher-order resonances. When we ignore the reflection phase, the FSR for different order resonances at $\lambda_0$ is

$$FSR_{m,m+1}^{\lambda_0} \sim \frac{\lambda_0}{m+1}.$$

In addition to FSR, multispectral imaging requires a narrow linewidth, which leads to more spectral channels and higher resolution. However, the FWHM of the resonant peaks is given by

$$FWHM_m^{\lambda_0} \sim \frac{2\lambda_0}{\pi m},$$

and increases for lower-order resonances, resulting in lower spectral resolution and fewer spectral channels. Previous work with first-order resonant peaks in MSFAs allows for a large FSR but with relatively large linewidth due to the properties of the lowest-order resonant peaks.[5,6,9,10,30] On the contrary, employing higher-order resonant peaks such as the second-order resonant peak, which decreases the FWHM by factor of 2, leads to a much smaller FSR, which limits the working wavelength range for the MSFA.[6,8,30–32]

We introduce a 10nm Pt layer to achieve a high FSR while maintaining a narrow linewidth. The Pt layer, which is thermally stable and CMOS compatible, sits in the middle of the cavity to suppress odd-order resonances, leading to a large FSR for the second-order resonant peak. The suppression enables the use of second-order resonances with an extended FSR, which breaks the traditional relationship between FSR and resonance order. The Pt layer can selectively suppress the resonant peak by taking advantage of the electric field distribution.[8] In the cavity, the resonance displays a clear intensity distribution. Specifically, the odd-order resonance has an electric field maximum at the middle of the cavity, while the electric field of even-order resonances is minimized in this area. By inserting a metal layer in the middle of cavity, the first transmission peak is the second-order resonant peak with the first and third-order resonant peaks suppressed (Figure 1c). The effective FSR of the second-order resonant peak is now the separation from the fourth-order resonant peak rather than the third-order resonant peak and is comparable to the FSR of the first-order resonant peak of a traditional Fabry–Pérot cavity.

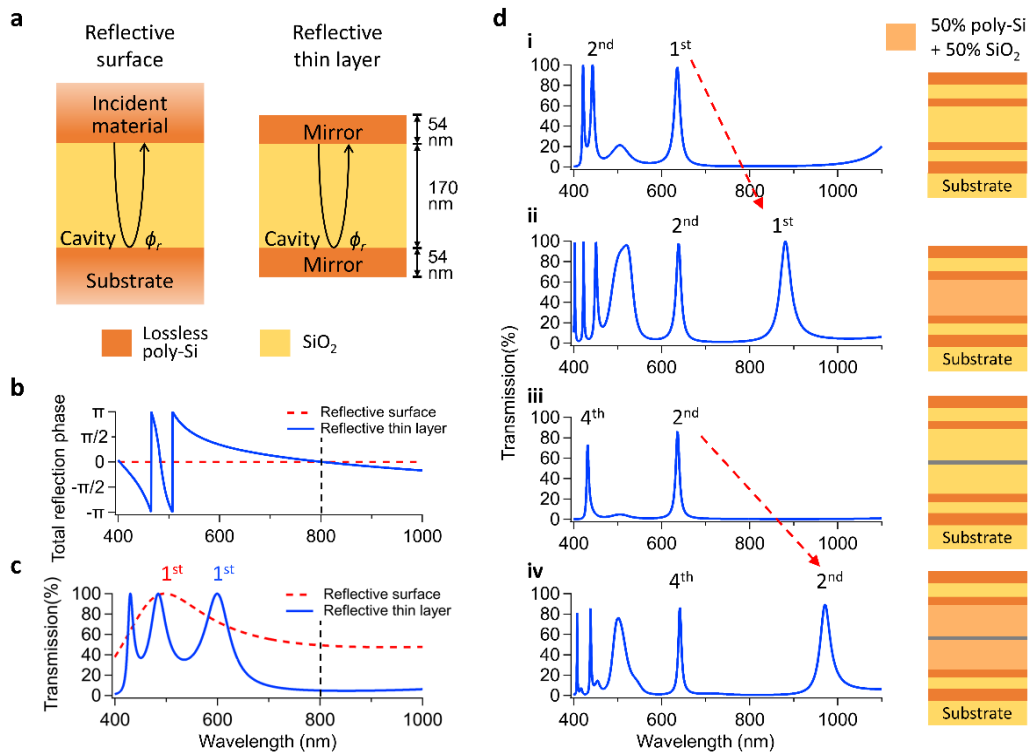

Figure 2. Illustration of the influence of reflection phase from the mirror on Fabry–Pérot cavities. The imaginary refractive index of polysilicon which is relatively small and minimally affects the reflection

phase and optical length is ignored for better illustration. (a) Schematic diagram illustrating two Fabry–Pérot cavities of equal thickness formed by either two reflective surfaces or two reflective thin layers. (b) Reflection phase from the reflective surfaces or reflective thin layers across different wavelengths. (c) Simulated transmission spectra for the two cases, highlighting the locations of the first-order resonant peaks. (d) A comparative analysis between designs utilizing the first-order resonant peak and the second-order resonant peak. Simulated spectra are shown for Fabry–Pérot structures with or without an inserted metal layer. (i) Cavity thickness of 212nm, with the first resonant peak at 635nm. (ii) Introduction of 50% lossless polysilicon into the cavity of (i) through a weighted averaging of permittivities. (iii) Cavity thickness of 370nm and 10nm pt layer in the middle, with the second resonant peak at 635nm. (iv) Introduction of 50% lossless polysilicon into the cavity of (iii) through a weighted averaging of permittivities.

When we consider the reflection phase ($\phi_r$) from the mirror part, the second-order resonance peak provides even larger FSR and tuning range, which surpasses that of the first-order resonance. The type of mirror used affects all three aspects - FSR, FWHM and transmission, of the Fabry–Pérot cavity. The mirror parts can be composed by metallic layers or distributed Bragg reflectors.[4,6,8–11,30–32] Metallic mirrors have a wide reflection band and a very small wavelength dependent reflection phase change due to the large imaginary refractive index and the small layer optical thickness compared to the working wavelength.[8–11,30] The weak reflection phase wavelength dependence makes the FSR depend mostly on the optical length of the cavity and not on the reflection phase. However, the large imaginary refractive index introduces a large absorption. A silver mirror is frequently used in Fabry–Pérot filters due to its high k and n ratio resulting in a relatively low loss compared with other metals. However, silver is not CMOS compatible and rules it out from monolithic integration into CMOS imaging sensors. Typically, a distributed Bragg mirror consists of pairs of high and low index layers with a quarter-wavelength thickness of the center wavelength of the working wavelength range.[4,6,31,32] The wavelength dependent reflection phase ($\phi_r$) causes both blue and red side resonant peaks to shift towards the center wavelength due to the sign change of $\phi_r$ at the center wavelength compared to the case $\phi_r = 0$ (Figure 2a-c). This effect becomes more significant as the peak moves farther from the center wavelength, which significantly reduces the FSR of resonant peaks and increases the optical length changes required for peak tuning. This effect is more pronounced in lower-order resonant peaks since the reflection phase takes up a larger proportion in the denominator of the resonance equation. Consequently, compared to the first-order resonant peak, the FSR and tuning range of the second resonant peak in our design can increase by more than 30% under the same cavity optical length change (Figure 2d). In our work, we use HLH three-layer thin films composed by polysilicon and silicon dioxide to form an asymmetric distributed Bragg reflector (two polysilicon layers with 40nm and 55nm sandwiching 120nm $SiO_2$). The asymmetric design used in each DBR is to extend the reflection bandwidth to wavelengths smaller than 600nm. This leads to a strong sideband suppression in the visible range by enhancing the absorption from polysilicon (Figure S2 in Supplementary information). Adding more pairs of polysilicon and silicon dioxide layers can further increase the reflection to get smaller FWHM. However, the reflection phase change will increase, which decreases the FSR and tuning range of second-order resonant peaks. Moreover, the use of the Pt layer also significantly suppresses side peaks and background transmission, resulting in enhanced color purity and contrast across the whole wavelength range (400nm-1000nm) of a general silicon photodetector.

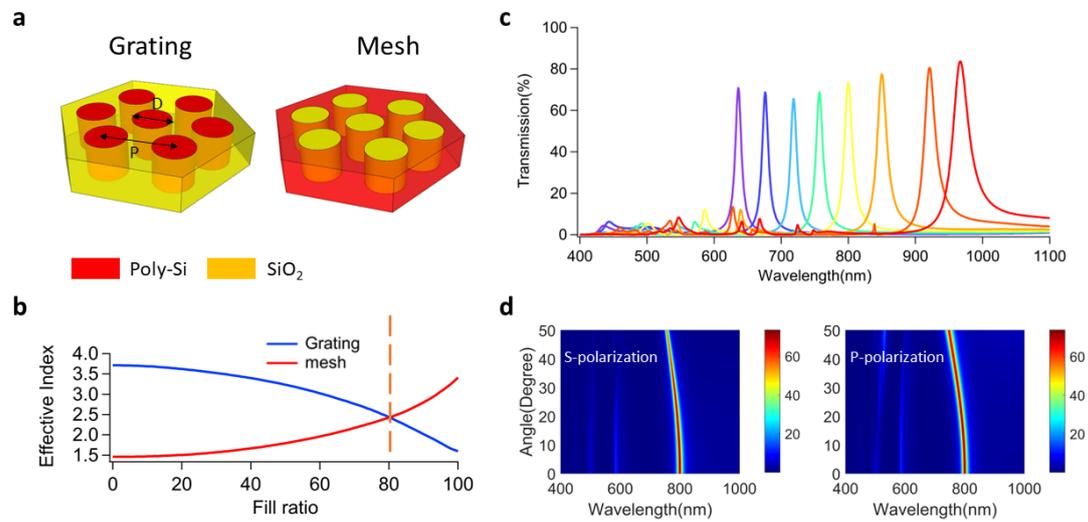

Figure 3. (a) Schematic diagram illustrating the grating and mesh subwavelength structures. (b) Calculated effective index for the grating and mesh structures with varying fill ratios (the ratio between the period (P) and diameter (D)). The period is fixed at 200nm, and the wavelength is set at 800nm. (c) Simulations of the transmission spectra for the designed filters. The filters with resonant peaks ranging from 630nm to 850nm adopt grating structures with diameters ranging from 0nm to 150nm and periods ranging from 160nm to 220nm. The filters with resonant peaks ranging from 850nm to 960nm utilize mesh structures with diameters ranging from 250nm to 350nm and periods ranging from 330nm to 400nm. (d) Simulation of the transmission spectra for the filter at 800nm with D=120nm and P=200nm under varying incident angles (0 to 50 degrees) and polarizations.

To maximize the wavelength tuning range in Fabry–Pérot color filter arrays, we employ both mesh and grating subwavelength structures in the cavity layer to modify the effective index while keeping feature sizes within the limits of advanced photolithography. According to effective-medium theory (EMT), the effective index of 2D subwavelength gratings is determined by the shape (local distribution) and area of the high and low index materials.[33–35] This means that equal area occupation of high or low index materials within the grating cell does not necessarily result in equal effective index. Two common distributions of subwavelength structures are the mesh and grating structures (Figure 3a).[35] In the grating structure, high-index nano-posts are surrounded by a low-index medium, which has an effective index closer to the low-index medium. On the other hand, a mesh structure consists of low-index regions surrounded by a high-index background, resulting in a high effective index (Figure 3b). Our multispectral filter arrays (MSFAs) employ the subwavelength grating structures for lower effective index parts and mesh structures for higher effective index parts to maximize the effective index range while minimizing the demands on the fabrication process. This allows for the entire index range to be obtained with fill ratios ranging from 0.2 to 0.8.

Here, we select polysilicon as the high index material due to its high refractive index and low absorption in the NIR range and $SiO_2$ as the low index material. To ensure the zero-order diffraction condition ( $period < \sim \frac{\lambda}{n}$ ) at the second-order resonant wavelength, we adjusted the period of the hexagonal lattice to range between 160nm to

350nm for different target bandpass wavelengths (see parameters combination in Table 1 in supplementary information). The subwavelength structures on both sides of Pt possess the same structures and physical thickness of 180nm, resulting in equal optical thickness. The side peaks from high-order diffraction at shorter wavelength are effectively suppressed by Pt and polysilicon which relaxes the subwavelength requirements to the wavelengths larger than the target wavelength, thereby easing the fabrication demands (Figure S3 in Supplementary information). The designed filter arrays are simulated with Rigorous Coupled-Wave Analysis (RCWA) and their transmission spectra are shown in Figure 3. The spectra cover a range from 630nm to 960nm, with a transmission larger than 60% and a FWHM ranging from 11nm to 29nm (Figure 3c). Low polarization and angular sensitivity are shown in Figure 3d. A high color purity across 400nm to 1000nm is obtained due to the absorption from the Pt layer and the slowly increasing polysilicon in the cavity.

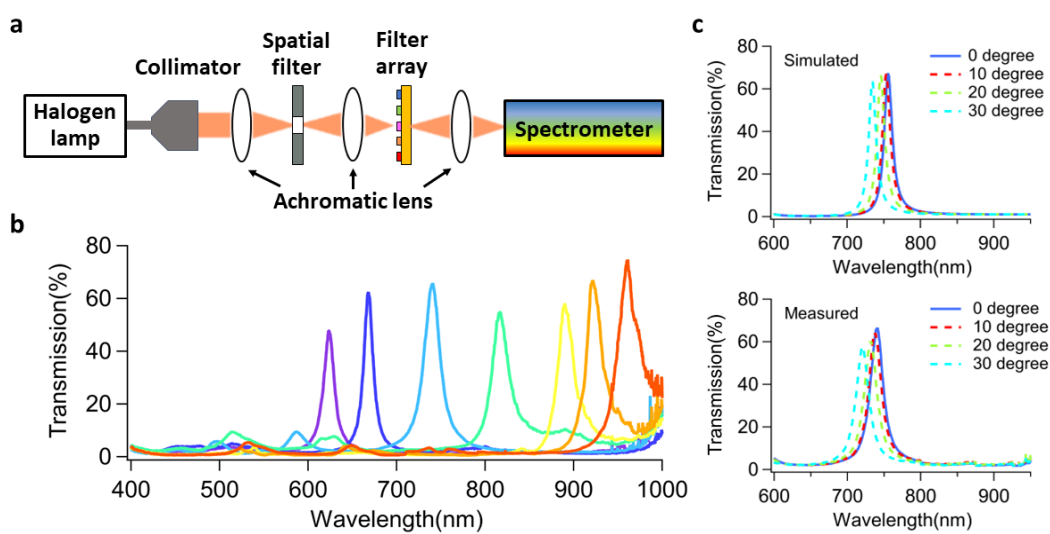

Figure 4. (a) Schematic diagram illustrating the measurement setup. (b) Transmission spectra of the designed filters obtained from measurements. (c) Simulated and measured transmission spectra for color filter at 750nm with D=110nm and P=220nm under different incident angles. For angles of ±20° (equivalent to F/# of 1.37) the filter spectrum is weakly affected, i.e., within its linewidth.

The optimized color filters with dimensions of 50μm x 50μm are fabricated using standard CMOS processes (see fabrication process in Figure S4 in Supplementary information). Polysilicon is obtained through furnace annealing of PECVD-deposited amorphous silicon, which can be replaced by laser annealing (to make it back-end CMOS compatible).[36,37] To ensure stability at high temperatures, platinum has been selected as the inserted metal layer, as most refractory metals tend to react with both $SiO_2$ and silicon under high temperatures.[38,39] To prevent the reaction between Pt and polysilicon during the annealing step, two 15nm silicon dioxide buffer layers sandwiching the Pt layer are employed (see Figure S5 in Supplementary information).[40] An antireflection layer and protection layer of 110nm $SiO_2$ are deposited above the color filter (Figure 1a). To accurately measure the transmission of the color filters, a 100nm Pt is deposited on top of the antireflection layer with a 40μm x 40μm aperture located at the color filters. The relative transmission is determined by comparing the output light through the color filters with that of a bare fused silica wafer with the same size aperture. The experimental setup used to measure the transmission spectra of the color filters is depicted in Figure 4a (see Methods section for further details). A broadband white light is focused on the filters, and the transmitted light is detected by a spectrometer. The measured transmission spectra for the fabricated color filter arrays are presented in Figure 4b.

The measured wavelength range of the color filter arrays extends from 622nm to 960nm, covering over 330nm from red to NIR. The measured FWHM ranges from 13nm to 31nm. The transmissions of all color filters are greater than 45% throughout the range with an average transmission larger than 60%. The sidebands are well suppressed across the whole range from 400nm to 1000nm which makes it compatible with most silicon CMOS imaging sensors without additional bandpass filters. Compared to the simulated results, the peak transmissions exhibit a decrease in transmission of 10-25%, which is attributed to the fabrication-induced asymmetry in the two cavities, which can be alleviated by further optimizing the fabrication process. The asymmetry in the two cavities causes a shift in the electric field distribution of the resonant peaks, resulting in increased absorption of the second-order resonant peak. However, this asymmetry enhances color purity by introducing more absorption for side peaks.

We demonstrate experimentally a weak dependence of the filter wavelength under oblique incidence. We investigate the behavior of our color filters under oblique incidence and measure the transmission spectra for incident angles ranging from 0 to 30 degrees. Figure 4c depicts the measured and simulated spectra for a color filter targeting 750nm under various incidence angles. We observe a shift of 10nm for incidence angles of 20°, equivalent to an F/# of 1.37 and a 20nm shift for incidence angles up to 30°. The simulations and measurement results are in good agreement, indicating that our color filters exhibit high angular insensitivity. In the absence of high-order diffraction, the shift in the transmission peak caused by oblique incidence is relatively small, as the refractive index of the cavity layer is higher than that of air.

Mixing a high index material like polysilicon in the cavity can further alleviate this effect, resulting in low crosstalk for side-by-side color filter arrays. Additionally, we study the effect of the filter size using finite difference time domain (FDTD) simulations, which demonstrates that the filters maintain their high performance even with a size smaller than 2μm, compatible with modern CMOS imagers (see Figure S6 in Supplementary information).

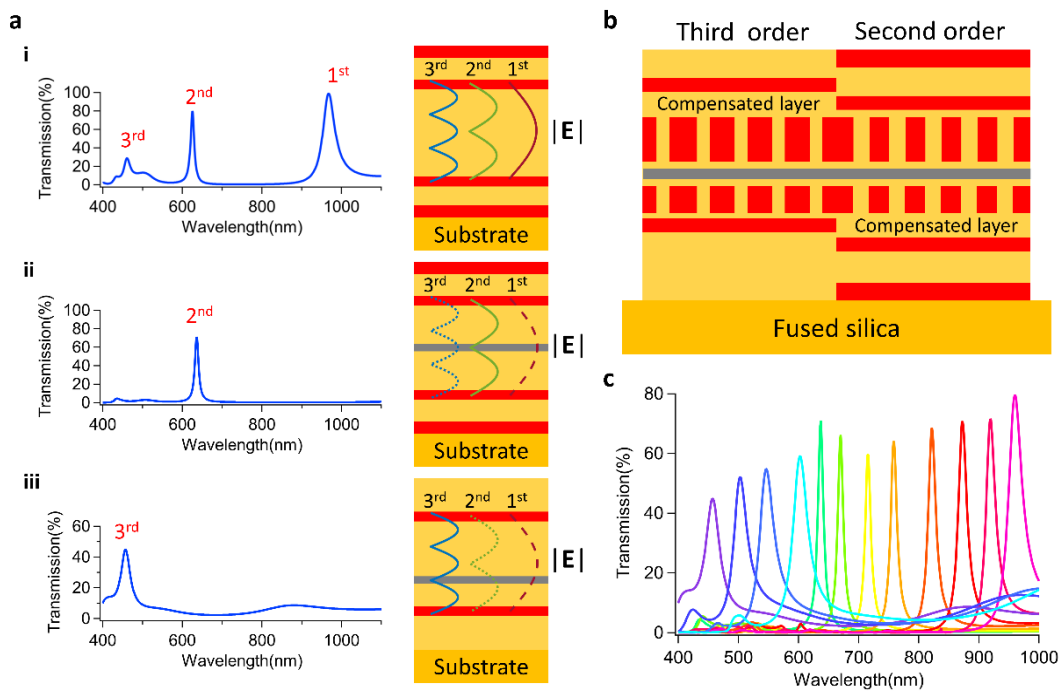

Figure 5. (a) Schematic diagram illustrating the utilization of the second or third-order resonances based on selective suppression. (i) Transmission spectra of previously demonstrated design without metal layer and polysilicon in the cavity. (ii) When the metal layer is located at the middle of cavity (the intensity minimum of second-order resonance). (iii) When the metal layer is located at the intensity minimum of third-order resonance (b) Schematic diagram of resonance combining design (c) Simulated Transmission spectra of resonance combining design.

We develop a new design strategy that combines second and third-order resonances to significantly expand the working wavelength range of our multispectral filter array to cover the full spectral window of silicon CMOS sensors. In our previous design, without a metal layer and polysilicon in the cavity, the first three resonant peaks are located as shown in Figure 5a (i). Notably, the third-order resonant peak is near the blue edge of silicon photodetectors at 460nm. By relocating the metal layer, we can switch to use the third-order resonance while effectively suppressing the first, second, and fourth-order resonances (Figure 5a (ii-iii)). To enhance transmission, we replace the 55nm polysilicon layer with silicon dioxide and similarly we can tune the transmission peaks by introducing subwavelength structures in the cavity. To combine the second and third-order resonances, we incorporate a compensating layer of silicon dioxide in each cavity to make the subwavelength structures of the two designs align at the same level. Thus, all the cavity layers are patterned in the same two high resolution lithography steps during fabrication (Figure 5b). With this resonance combining design, our MSFAs can cover a wide wavelength range from visible to near-infrared (455nm-960nm), achieving transmission greater than 40% and a full width at half maximum (FWHM) smaller than 32nm based on RCWA simulations (Figure 5c) (see parameters combination in Table 2 in Supplementary information). This extended range aligns with the silicon photodetection range, maximizing the working wavelength range for both filters and sensors.

In conclusion, we theoretically and experimentally demonstrate broadband multispectral color filter arrays based on a metal separated dual-cavity Fabry–Pérot architecture with subwavelength structures and present a resonance combining design to extend the working range over a full octave from 455nm to 960nm. In contrast to conventional Fabry–Pérot based multispectral filter arrays, our approach combines subwavelength structures with selective suppression to overcome the trade-off between the FSR and FWHM. By exploiting the reflection phase effect, we demonstrate that high-order resonances offer a larger tuning range with the same percentage of optical length change when compared to first order resonances. The resonance combining design utilizing both second and third-order resonances extends the wavelength range of our MSFAs to the whole detection range of most silicon photodetectors and introduces the possibility of using and combining different resonances in a single monolithic structure. Without relying on angle-sensitive diffraction structures, the color filters we developed are relatively insensitive to angular changes up to 30 degrees. Our investigation of the size effect suggests that our color filter arrays are highly customizable and compatible with current miniatured pixel sizes with a few microns. This finding makes it possible to directly replace conventional Bayer color filters, enabling the realization of a monolithically integrated multispectral CMOS sensor with a broad working wavelength range, high imaging contrast, and quality. The whole fabrication process is CMOS compatible. To make the fabrication of the filters compatible with the CMOS backend process, a potential alternative to the high temperature annealing method is the use of pulsed laser annealing, which is an industry standard.[36,37] The design principles demonstrated in this work can be easily transferred to other wavelength ranges such as near-infrared (NIR) and infrared (IR) by re-optimizing the thin film thicknesses and materials. At longer wavelengths, this approach can offer a larger wavelength range with easier fabrication than we demonstrated. Additionally, by exploring the freedom in the metal location, cavity and distributed Bragg reflector (DBR) phase manipulation, and introducing tunable materials, this design principle can be utilized not only for multispectral color filter arrays but also for more customized and even actively tunable color filters with various applications in imaging and display technologies.

Methods.

*Sample Fabrication*. The amorphous and silicon dioxide thin films are deposited on a fused silica substrate via Plasma-enhanced chemical vapor deposition (PECVD). Subsequently, the amorphous silicon is annealed to polysilicon in a furnace at 700 ˚C with Argon environment. The complex index data of the materials used are measured and characterized by a Woollam RC2 ellipsometer. The subwavelength structure is defined on the polysilicon surface by electron beam lithography methods using a negative resist, Hydrogen Silsesquioxane (HSQ). The patterns are transferred through inductively coupled plasma (ICP) etching. Hydrogen bromide (HBr) gas is used for ICP etching of polysilicon, which has a high selectivity between HSQ and

Polysilicon, enabling the definition of structures with high aspect ratios. The residual resist is removed using Hydrofluoric acid (HF). Then, a 400nm tetraethoxysilane (TEOS) oxide is deposited by PECVD to fill the gap in the polysilicon cavity, and the redundant oxide is polished by a chemical mechanical polishing (CMP) process. A 15nm oxide layer is deposited after the CMP process to act as a buffer layer, followed by 10nm Pt that is sputtered on the buffer layer. The top part of the filter is fabricated using similar processes.

*Measurement Procedure.* The color filters are fabricated to dimensions of 50μm by 50μm. A halogen lamp (HL-2000-LL) is used to generate a broadband light source. The collimated white light is focused on a spatial filter (circular pinhole with 50μm diameter) by an achromatic lens. The light after the spatial filter is refocused on the color filter with half angle smaller than 5°. The transmitted light is collected by another achromatic lens and detected with a spectrometer (Ocean Insight FLAME-S-VIS-NIR-ES). The sample is mounted on an XYZ stage to measure the transmission spectra for different color filters. The XYZ stage is mounted on a rotating breadboard to measure the transmission spectra for different incident angles. By comparing the transmission of filters with a reference sample, the relative transmission is acquired.

Acknowledgements

This work was performed in part at the Cornell NanoScale Facility, a member of the National Nanotechnology Coordinated Infrastructure (NNCI), which is supported by the National Science Foundation (Grant NNCI-2025233).

Reference

1. Lapray, P.-J., Wang, X., Thomas, J.-B. & Gouton, P. Multispectral Filter Arrays: Recent Advances and Practical Implementation. *Sensors* **14**, 21626–21659 (2014).

2. Qin, J., Chao, K., Kim, M. S., Lu, R. & Burks, T. F. Hyperspectral and multispectral imaging for evaluating food safety and quality. *J. Food Eng.* **118**, 157–171 (2013).

3. Hagen, N. A. & Kudenov, M. W. Review of snapshot spectral imaging technologies. *Opt. Eng.* **52**, 090901 (2013).

4. Kai Xu *et al.* All-Dielectric Color Filter with Ultra-Narrowed Linewidth. *Micromachines* **12**, 241 (2021).

5. Xiao, J., Song, F., Han, K. & Seo, S.-W. Fabrication of CMOS-compatible optical filter arrays using gray-scale lithography. *J. Micromechanics Microengineering* **22**, 025006 (2012).

6. Williams, C., Gordon, G. S. D., Wilkinson, T. D. & Bohndiek, S. E. Grayscale-to-Color: Scalable Fabrication of Custom Multispectral Filter Arrays. *ACS Photonics* **6**, 3132–3141 (2019).

7. Shen, Y., Istock, A., Zaman, A., Woidt, C. & Hillmer, H. Fabrication and characterization of multi-stopband Fabry–Pérot filter array for nanospectrometers in the VIS range using SCIL nanoimprint technology. *Appl. Nanosci.* **8**, 1415–1425 (2018).

8. Lee, K.-T., Han, S. Y., Li, Z., Baac, H. W. & Park, H. J. Flexible High-Color-Purity Structural Color Filters Based on a Higher-Order Optical Resonance Suppression. *Sci. Rep.* **9**, 14917 (2019).

9. Park, C.-S., Shrestha, V. R., Lee, S.-S. & Choi, D.-Y. Trans-Reflective Color Filters Based on a Phase Compensated Etalon Enabling Adjustable Color Saturation. *Sci. Rep.* **6**, 25496 (2016).

10. Mao, K. *et al.* Angle Insensitive Color Filters in Transmission Covering the Visible Region. *Sci. Rep.* **6**, 19289 (2016).

11. Yoon, Y.-T. & Lee, S.-S. Transmission type color filter incorporating a silver film based etalon. *Opt. Express* **18**, 5344 (2010).

12. Sakat, E. *et al.* Free-standing guided-mode resonance band-pass filters: from 1D to 2D structures. *Opt. Express* **20**, 13082 (2012).

13. Kaplan, A. F., Xu, T. & Jay Guo, L. High efficiency resonance-based spectrum filters with tunable transmission bandwidth fabricated using nanoimprint lithography. *Appl. Phys. Lett.* **99**, 143111 (2011).

14. Park, C.-H., Yoon, Y.-T. & Lee, S.-S. Polarization-independent visible wavelength filter incorporating a symmetric metal-dielectric resonant structure. *Opt. Express* **20**, 23769 (2012).

15. Mazulquim, D. B. *et al.* Efficient band-pass color filters enabled by resonant modes and plasmons near the Rayleigh anomaly. *Opt. Express* **22**, 30843 (2014).

16. Dong, Z. *et al.* Printing Beyond sRGB Color Gamut by Mimicking Silicon Nanostructures in Free-Space. *Nano Lett.* **17**, 7620–7628 (2017).

17. Park, H. & Crozier, K. B. Multispectral imaging with vertical silicon nanowires. *Sci. Rep.* **3**, 2460 (2013).

18. Horie, Y. *et al.* Visible Wavelength Color Filters Using Dielectric Subwavelength

Gratings for Backside-Illuminated CMOS Image Sensor Technologies. *Nano Lett.* **17**, 3159–3164 (2017).

19. Horie, Y., Arbabi, A., Han, S. & Faraon, A. High resolution on-chip optical filter array based on double subwavelength grating reflectors. *Opt. Express* **23**, 29848 (2015).

20. Xu, T., Wu, Y.-K., Luo, X. & Guo, L. J. Plasmonic nanoresonators for high-resolution colour filtering and spectral imaging. *Nat. Commun.* **1**, 59 (2010).

21. Fleischman, D., Sweatlock, L. A., Murakami, H. & Atwater, H. Hyper-selective plasmonic color filters. *Opt. Express* **25**, 27386 (2017).

22. He, X. *et al.* A single sensor based multispectral imaging camera using a narrow spectral band color mosaic integrated on the monochrome CMOS image sensor. *APL Photonics* **5**, 046104 (2020).

23. Rajasekharan, R. *et al.* Filling schemes at submicron scale: Development of submicron sized plasmonic colour filters. *Sci. Rep.* **4**, 6435 (2015).

24. Chen, Q. & Cumming, D. R. S. High transmission and low color cross-talk plasmonic color filters using triangular-lattice hole arrays in aluminum films. *Opt. Express* **18**, 14056 (2010).

25. Shah, Y. D. *et al.* Ultralow-light-level color image reconstruction using high-efficiency plasmonic metasurface mosaic filters. *Optica* **7**, 632 (2020).

26. Yokogawa, S., Burgos, S. P. & Atwater, H. A. Plasmonic Color Filters for CMOS Image Sensor Applications. *Nano Lett.* **12**, 4349–4354 (2012).

27. Miyamichi, A., Ono, A., Kagawa, K., Yasutomi, K. & Kawahito, S. Plasmonic

Color Filter Array with High Color Purity for CMOS Image Sensors. *Sensors* **19**, 1750 (2019).

28. Fleischman, D. *et al.* High Spectral Resolution Plasmonic Color Filters with Subwavelength Dimensions. *ACS Photonics* **6**, 332–338 (2019).

29. Shrestha, V. R., Park, C.-S. & Lee, S.-S. Enhancement of color saturation and color gamut enabled by a dual-band color filter exhibiting an adjustable spectral response. *Opt. Express* **22**, 3691–3704 (2014).

30. Walls, K. *et al.* Narrowband multispectral filter set for visible band. *Opt. Express* **20**, 21917–21923 (2012).

31. McClung, A., Samudrala, S., Torfeh, M., Mansouree, M. & Arbabi, A. Snapshot spectral imaging with parallel metasystems. *Sci. Adv.* **6**, eabc7646 (2020).

32. Horie, Y., Arbabi, A., Arbabi, E., Kamali, S. M. & Faraon, A. Wide bandwidth and high resolution planar filter array based on DBR-metasurface-DBR structures. *Opt. Express* **24**, 11677 (2016).

33. Kikuta, H., Ohira, Y., Kubo, H. & Iwata, K. Effective medium theory of two-dimensional subwavelength gratings in the non-quasi-static limit. *J. Opt. Soc. Am. A* **15**, 1577 (1998).

34. Grann, E. B., Moharam, M. G. & Pommet, D. A. Artificial uniaxial and biaxial dielectrics with use of two-dimensional subwavelength binary gratings. *J. Opt. Soc. Am. A* **11**, 2695 (1994).

35. Grann, E. B. & Moharam, M. G. Hybrid two-dimensional subwavelength surface-relief grating–mesh structures. *Appl. Opt.* **35**, 795 (1996).

36. Bronnikov, K. *et al.* Large-Scale and Localized Laser Crystallization of Optically Thick Amorphous Silicon Films by Near-IR Femtosecond Pulses. *Materials* **13**, 5296 (2020).

37. Parr, A. A. *et al.* A comparison of laser- and furnace-annealed polysilicon structure. *Semicond. Sci. Technol.* **17**, 47–54 (2002).

38. Chen, W. D., Cui, Y. D., Hsu, C. C. & Tao, J. Interaction of Co with Si and $SiO_2$ during rapid thermal annealing. *J. Appl. Phys.* **69**, 7612–7619 (1991).

39. Dadabhai, F., Gaspari, F., Zukotynski, S. & Bland, C. Reduction of silicon dioxide by aluminum in metal–oxide–semiconductor structures. *J. Appl. Phys.* **80**, 6505–6509 (1996).

40. Shi, J., Kojima, D. & Hashimoto, M. The interaction between platinum films and silicon substrates: Effects of substrate bias during sputtering deposition. *J. Appl. Phys.* **88**, 1679–1683 (2000).